\RequirePackage{fix-cm}

\documentclass[conference]{IEEEtran}

\IEEEoverridecommandlockouts

\usepackage{algorithmic}
\usepackage[compress]{cite}
\usepackage{graphicx}
\usepackage{amsmath,amssymb,amsfonts}
\usepackage{textcomp}
\usepackage{xcolor}
\usepackage{subfigure}
\usepackage{xspace}
\usepackage{mathtools}
\usepackage{enumitem}
\usepackage{booktabs}
\usepackage{multirow}

\usepackage{siunitx}
\usepackage{verbatim}
\usepackage{mdframed} 
\usepackage{listings}

\usepackage{url}
\usepackage[hidelinks]{hyperref}
\usepackage{multirow}

\def\BibTeX{{\rm B\kern-.05em{\sc i\kern-.025em b}\kern-.08em
    T\kern-.1667em\lower.7ex\hbox{E}\kern-.125emX}}
\begin{document}

\title{JupOtter: Cell-Level Bug Detection in Jupyter Notebooks
\thanks{This work was supported by the Natural Sciences and Engineering Research Council of Canada (NSERC).}
}

\author{\IEEEauthorblockN{Lukas Ottenhof}
\IEEEauthorblockA{
\textit{University of Alberta}\\
Edmonton, AB, Canada \\
lottenho@ualberta.ca}
\and
\IEEEauthorblockN{Thibaud Lutellier}
\IEEEauthorblockA{
\textit{University of Alberta}\\
Camrose, AB, Canada \\
lutellie@ualberta.ca}
}


\definecolor{emerald}{rgb}{0.31, 0.78, 0.47}
\newcommand{\rev}[1]{{\color{red}#1}}
\newcommand{\todoc}[2]{{\textcolor{#1}{\textbf{#2}}}}
\newcommand{\todoblack}[1]{{\todoc{black}{\textbf{[[#1]]}}}}
\newcommand{\todored}[1]{{\todoc{red}{\textbf{[[#1]]}}}}
\newcommand{\todogreen}[1]{\todoc{emerald}{\textbf{[[#1]]}}}
\newcommand{\todoblue}[1]{\todoc{blue}{\textbf{[[#1]]}}}
\newcommand{\todoorange}[1]{\todoc{orange}{\textbf{[[#1]]}}}
\newcommand{\todobrown}[1]{\todoc{brown}{\textbf{[[#1]]}}}
\newcommand{\todogray}[1]{\todoc{gray}{\textbf{[[#1]]}}}
\newcommand{\todopurple}[1]{\todoc{purple}{\textbf{[[#1]]}}}
\newcommand{\todopink}[1]{\todoc{magenta}{\textbf{[[#1]]}}}
\newcommand{\todocyan}[1]{\todoc{cyan}{\textbf{[[#1]]}}}
\newcommand{\todoviolet}[1]{\todoc{violet}{\textbf{[[#1]]}}}
\newcommand{\todo}[1]{\todoorange{TODO: #1}}

\newcommand{\thibaud}[1]{\todoblue{Thibaud: #1}}

\newcommand{\lukas}[1]{\todogreen{Lukas: #1}}

\newcommand{\code}[1]{\texttt{\small #1}} 


\newcounter{finding}
\newcommand{\finding}[1]{\refstepcounter{finding}
	\begin{mdframed}[linecolor=gray,roundcorner=12pt,backgroundcolor=gray!15,linewidth=3pt,innerleftmargin=2pt, leftmargin=0cm,rightmargin=0cm,topline=false,bottomline=false,rightline=false]
	 #1
      \end{mdframed}
}

\maketitle
\begin{abstract}
Jupyter Notebooks are an increasingly popular coding environment used across many domains, especially in Python-based data science and scientific computing. 
Originally used for prototyping and interactive exploration, notebooks are increasingly used to develop more complex programs, leading to a rapid rise in buggy notebooks on platforms like GitHub. 
To address this trend, we present \textbf{JupOtter}, a bug detection system designed specifically for Jupyter Notebooks. 
JupOtter features three novel contributions: (1) a notebook-specific tokenization strategy that preserves cell structure, (2) a cell-level bug prediction technique, and (3) a new labeled dataset, \textbf{OtterDataset}, containing over 21,000 notebooks annotated for fine-grained cell-level bug detection.
JupOtter achieves cell-level bug detection F1 scores that surpass static analyzers and large language models in two out of three evaluation datasets.
\end{abstract}


\section{Introduction}
\label{sec:intro}


Jupyter Notebooks have become the standard environment in data science due to their unique interactive structure~\cite{perkel2018jupyter}. 
Unlike traditional code files, notebooks consist of executable cells that can contain either code or markdown, and can be executed in any order. 
This cell-based execution model allows users to run time-consuming operations (e.g., data loading) once and iteratively modify downstream code without re-executing the entire notebook. 
While this flexibility has accelerated development and experimentation, it has also introduced new challenges in code reliability. As notebooks evolve from prototyping tools into complex, multi-stage software artifacts, the risk of implementation bugs grows. This shift has contributed to a rising number of buggy notebooks~\cite{de2024bug}.


Recent studies show that the number of implementation bugs in Jupyter Notebooks found on Stack Overflow is increasing by 48\% annually~\cite{de2024bug}, suggesting developers are struggling to manage notebook quality. This problem is especially concerning in domains such as scientific research, finance, and artificial intelligence, where high accuracy and reliability are critical. This paper presents contributions focused on detecting bugs related to code implementation because they represent a significant and growing portion of real-world notebook bugs. Unlike environment-related bugs that depend on external dependencies or configurations, implementation bugs are self-contained within notebooks and can be identified by learning patterns in code. Detecting such bugs prior to execution can reduce debugging time and resource costs.

Jupyter Notebooks present unique challenges for bug detection due to their non-linear, cell-based execution model. Unlike traditional scripts, notebooks allow cells to be run in arbitrary order, which can introduce implementation bugs tied to execution state and cell dependencies. As a result, conventional debugging techniques that work on sequential scripts are often ineffective. 

Machine learning tools have shown potential in code understanding tasks such as bug detection across programming languages~\cite{allamanis2021self, wang2023defecthunter, li2024if, hossain2024deep,pradel2018deepbugs}. 
However, many existing machine learning-based bug detectors operate at the file or function level, which is ill-suited to the notebook format.
Simply knowing that a bug exists in a notebook file is often not actionable, and focusing on a single function may miss cross-cell dependencies. 
Despite recent advances, existing machine learning tools lack support for the structural and execution-specific characteristics of Jupyter Notebooks. 
As a result, the software engineering community still lacks effective approaches to automatically detect notebook bugs~\cite{wang2020better,jiang2025exploring}.  


To address the unique challenges of analyzing Jupyter Notebooks, we propose \textbf{JupOtter}: a novel machine learning-based bug detection system tailored to the notebook format. JupOtter introduces three key contributions: 

\smallskip\noindent(1) A new \textbf{tokenization strategy} which enables fine-grained predictions by inserting special tokens that wrap each cell and act as explicit cell boundaries. These boundary tokens allow models to know exactly where notebook cells start and end after tokenization, which is necessary for making precise, localized predictions. Unlike prior code tokenization approaches, which either flatten notebooks into plain scripts or truncate long inputs~\cite{feng2020codebert, wang2021codet5, guo2020graphcodebert, guo2022unixcoder}, our strategy preserves the original cell structure while enabling long-context processing without losing global notebook coverage. This makes fine-grained, cell-level reasoning possible at scale.

\smallskip\noindent(2) A new \textbf{cell-level bug prediction technique} which allows models to make separate predictions for each cell in a notebook. Unlike previous models built for linear, traditional source code~\cite{pornprasit2022deeplinedp, li2017software, dam2018deep, hin2022linevd}, our technique assumes no fixed execution order. This matches the naturally non-linear, cell-by-cell execution of computational notebooks. This differs from standard pretrained-model adaptation for sequence classification, which produces a single label per input sequence rather than per-cell representations.

\smallskip\noindent(3) OtterDataset, a \textbf{new dataset} of 21,303 Python Jupyter Notebooks labeled for bug detection. This dataset is designed to support cell-level and file-level evaluation and training, while reflecting real-world notebook usage. To the best of our knowledge, this is the first publicly available large-scale dataset of Jupyter Notebooks with cell-level labels.

\smallskip
JupOtter was evaluated using three datasets excluded from its training data (OtterDataset, Jupyter Errors~\cite{grotov2024untangling}, and CodeParrot~\cite{codeparrotgithubjupyter}). We aimed to answer four research questions: \textbf{RQ1:} How well does JupOtter detect bugs at the cell level, \textbf{RQ2:} How does JupOtter perform when detecting specific error types, \textbf{RQ3:} How does JupOtter compare to related tools for file-level bug detection, and \textbf{RQ4:} How does each contribution affect performance?

\section{Approach}
\label{sec:approach}

\begin{figure*}[ht]
    \centering
    \includegraphics[width=0.9\textwidth]{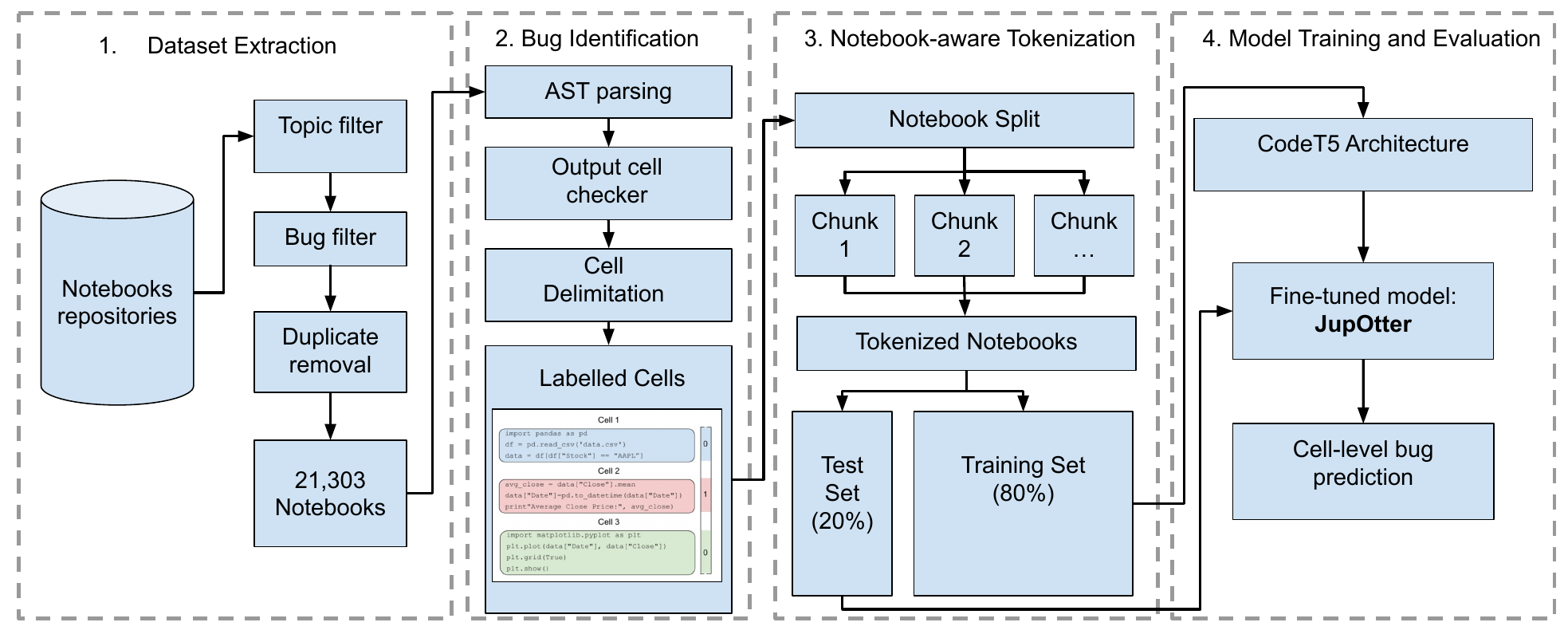}
    \caption{Overview of JupOtter, a bug detection system for Jupyter Notebooks. Our approach begins with source notebooks whose code cells are first extracted and preprocessed. The processed code cells are then tokenized into non-overlapping chunks using our tokenization strategy. These chunks are then processed by a pre-trained encoder, and predictions are generated.}
    \label{fig:overview}
\end{figure*}

At a high level, JupOtter processes Jupyter Notebooks by extracting code cells, converting them into a tokenized format that preserves notebook structure, and predicting whether each cell contains an implementation bug. Our approach is designed to handle the unique characteristics of notebooks, including variable-length cells and interleaved code and markdown content. Our approach consists of four main components: (1) a dataset construction phase, (2) a notebook-specific tokenization strategy, (3) a cell-level defect prediction model, and (4) a multi-segment training strategy (see Figure~\ref{fig:overview}).

\subsection{The OtterDataset}
\label{sub:extraction}
To address the lack of publicly available datasets for supervised cell-level bug detection in Jupyter Notebooks, we constructed OtterDataset: a new labeled dataset of Python notebooks designed for this task.
We systematically extracted Jupyter Notebook files from GitHub using GitHub Search API and a set of 32 search queries to ensure broad coverage of topics and error types. 
After cleaning, labeling, and removing duplicate notebooks, this process resulted in 9,937 buggy notebooks and 11,366 non-buggy notebooks (47\% buggy ratio, 21,303 total notebooks). At the cell-level, 43,789 cells are buggy out of 661,909 cells (7\% buggy cell ratio), with an average
of 2.1 buggy cells per notebook. To prevent over-representation of any particular query, we limited the collection to a maximum of 1,000 notebooks per query. 

The search queries were designed to capture a broad range of domains such as statistics and machine learning, and common bug types. To target common implementation errors, we incorporated bug-related keywords including Python exceptions such as ``IndexError'' and ``TypeError'', based on the list of built-in exceptions in the official Python documentation~\cite{PythonExceptions}. When constructing OtterDataset, we targeted environment-related errors and implementation-related errors since they are the most common categories of errors in notebooks~\cite{de2024bug}. 
We provided the complete list of queries used in our supplementary material, along with the number of notebooks they contributed to OtterDataset~\cite{jupotter_repo}.

Each notebook collected was assigned a unique identifier based on its author name, repository name, and file name to trace its origin. We used this identifier to remove duplicate notebooks that appeared in different queries to limit data leakage in our train-test split. We further validated separation between training and test sets at the author-repository level. 83.9\% of test notebooks in OtterDataset come from author-repository combinations not present in the training set.

To assess whether OtterDataset contains notebooks intentionally designed to demonstrate bugs (which would not be representative of real bugs), we randomly sampled and manually inspected 100 notebooks. None were explicitly focused on teaching or demonstrating specific bugs. While some notebooks were educational (e.g., demonstrating library usage), they did not intentionally inject errors. Although such bug-focused notebooks may exist, this inspection suggests they do not represent a substantial portion of the dataset.

\subsubsection{Notebook Bug Identification}
\label{subsub:notebook error annotation}

Previous studies have shown that reproducing notebook executions at scale is often unreliable due to missing dependencies, non-trivial environments, and non-linear cell execution~\cite{pimentel2019large}. To avoid these issues when creating labels, we did not attempt to re-execute the notebooks. Instead, we analyzed the output field of each executed code cell already present in the notebooks. Prior work has used stored cell outputs to detect and analyze bugs or errors in notebooks~\cite{wang2025machine, grotov2024untangling}, and has shown that over 85\% of notebooks contain executed cells with output data~\cite{pimentel2019large}. From these saved outputs, we extracted and labeled all bug-related information, including the error type, the cell and line number where the error occurred, and the error message. This information was used to label each cell as containing a bug or not, as well as to record the type of bug encountered.

After analyzing all saved cell outputs, we parsed the code cells of each notebook to identify syntax errors that might not be visible in notebook outputs (e.g., if a cell was never executed). Each code cell was parsed individually using Python’s \code{ast} module, which raises an exception when a syntax error is encountered. We parsed each cell in addition to examining saved cell outputs as a data cleaning step, to avoid having trivial syntax errors incorrectly labeled as correct code.

To handle notebook-specific commands such as \code{\%magic} and \code{!shell}, we first transformed each cell using IPython’s \code{TransformerManager}. This preprocessing step converts Jupyter-specific syntax into valid Python, ensuring that such cells are not incorrectly flagged as buggy by the parser.

When constructing OtterDataset, we targeted a wide variety of bug types to create a dataset representative of notebooks found online. However, during our labeling process, we excluded errors caused by the execution environment rather than the notebook code: \code{ImportError}, \code{FileNotFoundError}, \code{KeyboardInterrupt}, \code{SystemExit}, and \code{ConnectionError}. These were ignored to avoid introducing spurious labels. For example, two notebooks with identical code might differ only due to a user interruption, which would otherwise be incorrectly flagged as a runtime error. Additionally, we considered excluding memory errors, but we still labeled them as they can also be caused by bugs in the code (e.g., memory leaks or poor implementation). Although error types and messages are stored in the dataset, they were not used during model training, only during evaluation to assess performance on specific bug types. Since the goal of JupOtter is to detect the buggy cell \textbf{prior to execution}, such information would not be available in a real-world scenario. This metadata is included solely to support dataset analysis, evaluation, and future research.

Based on the identified errors, we generated a binary label list for each notebook, where each element corresponds to a code cell. A value of 1 indicates that the cell contains a bug, and 0 otherwise. Additionally, we assigned a file-level label marked as buggy if at least one of its code cells contained a bug. These labels were created to enable supervised learning at the file and cell-level and to address the lack of publicly available datasets for bug detection in Jupyter Notebooks.

To validate the accuracy of our labeling process, we randomly selected 30 notebooks from OtterDataset for manual label verification. 14 of the sampled notebooks contained at least one cell that could not be executed due to missing dependencies or unavailable data. For these cells, labels were assigned through manual inspection of the code, and when possible, they were executed using dummy input data to confirm expected behavior. These dependency issues differ from the excluded environment categories because the sampled notebooks contain cells whose dependencies were unavailable at verification time, but their bugs when present, are implementation-related.

24 of the 30 notebooks had identical manual and automated labels. In total, these notebooks contained 970 cells, of which 22 (2.3\%) of the automated labels differed from their manual counterparts. With a sample size of 970 cells, this corresponds to a margin of error of $\pm 0.94\%$ at a 95\% confidence level (Wald interval). 

We compared the manual and automated labels using a bootstrap McNemar test with 10,000 resamples and a sample size of 100 to assess statistical agreement. Out of the 10,000 resamples, only 139 (1.39\%) were significant at $p < 0.05$, indicating strong agreement between the automated and manual labels. The median p-value was 0.5000, with a 5th percentile of 0.1250 and a 95th percentile of 1.0000. These results suggest that the automated labeling procedure is largely consistent with manually verified labels, providing confidence in its reliability for large-scale notebook analysis.

\subsubsection{Dataset Overview}

Each notebook entry in OtterDataset includes the following seven features:

\begin{itemize}
    \item \textbf{notebook\_name:} A unique identifier composed of the author's username, repository, and file name.
    \item \textbf{notebook\_structure:} Concatenated code cells only, with cell boundary markers.

    \item \textbf{is\_buggy:} A binary label indicating whether the notebook contains at least one bug.
    \item \textbf{error\_locations:} A list of cell and line numbers where errors occurred.
    \item \textbf{error\_type:} The type(s) of detected error(s).

    \item \textbf{error\_message:} error message(s) detected.
    \item \textbf{buggy\_cells:} A binary list containing a label for each cell in a given notebook (e.g., \verb|[1, 0, 0, 1]|).
\end{itemize}

Since a notebook can contain multiple errors (e.g., an indentation error in cell 1 and an index error in cell 7), we tracked all errors using lists for both error types and error messages. These are linked through the \texttt{error\_locations} column, which combines information from multiple fields to give a detailed description of each buggy cell, including the error type, cell number, and specific line.

\subsection{JupOtter Defect Prediction System}
\label{sub:aproach_contributions}

In this section, we introduce three key contributions designed specifically for cell-level bug detection in Jupyter Notebooks: 
(1) a notebook-aware tokenization strategy that preserves cell boundaries and efficiently handles large notebook files; 
(2) a novel prediction method that identifies individual buggy cells within notebooks; and
(3) a mixed-precision, per-sample training loop that accommodates the multi-segment structure of tokenized notebook content. 
Together, these contributions form JupOtter, a cell-level bug detection system for Jupyter Notebooks.

\subsubsection{Notebook-aware tokenization strategy}

To tokenize input for our cell-level prediction model, we developed a chunking-based strategy that segments notebooks into multiple non-overlapping sequences of cells. 
This method builds upon prior work that handles large inputs by treating files as non-overlapping pages~\cite{liu2022leveraging}. Our approach preserves notebook structure and content without truncation and represents each notebook with three two-dimen\-sional tensors: one for input tokens, one for attention masks, and one for labels. Each row of these tensors contains a chunk of notebook cells whose combined token length does not exceed a user-defined limit. Additionally, we insert special tokens to mark cell boundaries within tokenized sequences, enabling precise identification of each cell’s position after tokenization. 

We divided notebooks into multiple chunks to ensure that notebooks exceeding the token limit of encoder models could still be fully tokenized and utilized for training without truncation.
Special tokens marking cell boundaries allow precise cell-level predictions directly from tokenized outputs. 
By enabling full processing of large notebooks, our strategy significantly enhances real-world applicability, removing constraints associated with processing only smaller notebooks.

Without our tokenization strategy, standard tokenization methods would truncate or split longer notebooks, discarding important syntactic and semantic information at the end of these sequences. This truncation would bias models towards shorter notebooks, limiting their effectiveness on realistic data. By training models with our strategy, we preserve the full context of larger notebooks, enabling models to learn from complete cell sequences and effectively handle realistic, large-scale notebook inputs.


To support cell-level predictions, it is critical to preserve cell boundaries during tokenization. We achieve this by adding special start and end cell markers. 
Each notebook is tokenized individually, processing cells sequentially and organizing them into chunks.
Cells are added to the current chunk until including another cell would exceed the model's input constraints. Once the limit is reached, the tokens and attention masks are padded to match the chunk length and stored as rows in their respective two-dimensional tensors. 
Corresponding labels for each chunk are recorded separately in a label tensor. 
The process continues iteratively until the entire notebook is tokenized. 
If a notebook’s total token length never exceeds the chunk length, it is stored as a single padded chunk comprising tensors for tokens, attention masks, and labels.

The only case where truncation occurs in our tokenization strategy is when a single cell exceeds the maximum chunk token length. 
Cells exceeding the chunk length are rare and resemble standalone scripts more than typical notebook cells, and are thus outside the intended scope of our model’s learning strategy. 987 of 661,909 cells in OtterDataset (0.15\%) exceed our chunk length of 2,500 tokens.

\subsubsection{Prediction Model's Architecture}

\begin{figure}[tbp]
    \centering
    \includegraphics[width=0.40\textwidth]{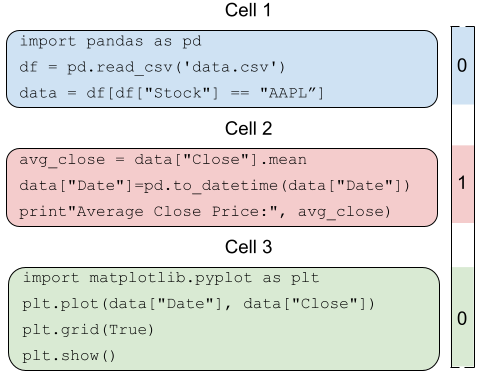}

    \caption{Example of a cell-level prediction output. Each cell is processed and assigned a binary label: 0 for non-buggy cells, 1 for buggy cells. }
    \label{fig:prediction_strat}
\end{figure}

Our approach is built on top of a pretrained transformer-based encoder (e.g., CodeBERT or CodeT5) which is well-suited for code understanding tasks. JupOtter predicts bugs at the cell-level by using boundary markers to identify cell boundaries within the model’s input. These markers allow the model to generate a binary prediction for each cell in a notebook chunk, indicating whether that cell is buggy (1) or not (0). Unlike traditional file-level models that produce a single label per notebook, our approach provides fine-grained predictions aligned with the notebook’s structure (Figure~\ref{fig:prediction_strat}).

File-level models cannot produce such localized predictions because they are architecturally constrained to aggregate information into a single global representation and output a single label per file. In contrast, our method explicitly preserves cell boundaries in the input representation and associates predictions with specific structural units of the notebook. Having cell-level predictions is more actionable and self-explanatory than a file-level prediction. For example, while informing a user that a file contains a bug provides guidance, identifying a buggy cell (mean lines of code per cell in OtterDataset is 6.43) provides more precise information and may allow users to take action without additional explanation.

To perform cell-level prediction, each tokenized chunk (which may contain several cells) is first passed through a transformer-based encoder to obtain contextualized hidden states. We then identify cell boundaries within the chunk using the special start and end markers and group the hidden states accordingly. For each cell, we compute a single representation by averaging the hidden state vectors of all tokens within its boundaries, including the boundary markers. This cell-level embedding is passed through a classifier (a linear layer) that outputs a single logit indicating the likelihood that the cell is buggy. Repeating this process for all cells in a chunk yields a vector of cell-level predictions. When processing large notebooks spanning multiple chunks, predictions from each chunk are concatenated to produce a final prediction.

Passing tokenized chunks through a transformer-based encoder prior to making predictions allows the model to capture inter-cell relationships without assuming a particular execution order. This helps address the challenge of non-linear execution in notebooks, avoiding false positives that dynamic tools may introduce when guessing execution order, and that static analysis tools may introduce when attempting to parse code sequentially. We investigate the impact of the encoder in our ablation study (RQ4).

We compute loss across notebook chunks using \code{BCEWithLogitsLo\-ss}, which combines a sigmoid activation with binary cross-entropy loss.
Logits and ground-truth labels from all chunks of a notebook are concatenated into a single prediction and label vector. 
Loss is then computed over these vectors, ensuring that each cell contributes equally to the notebook's total loss.

\subsubsection{Training strategy}

Our training strategy is designed to effectively handle large notebooks represented as multi-chunk inputs produced by our tokenization strategy. To accommodate this structure, we implemented a training loop that processes the chunks of each notebook individually within each batch. 

During each epoch, we iterate through batches of notebooks using a user-defined batch size. During training, we use mixed precision training with \code{torch.cuda.amp.autocast} and \code{GradScaler}.

In training, each sample's loss is scaled by dividing by batch size to normalize gradient contributions across the batch. 
We perform the backward pass on each sample’s scaled loss individually, accumulating gradients across the entire batch. This per-sample backward pass strategy is necessary due to memory constraints caused by large notebooks, and is repeated until all samples in the batch have been processed, accumulating loss and gradients across samples. 
After all samples in a batch are processed, we perform a single optimizer step and gradient scaler update using the accumulated gradients from the per-sample backward passes.

To track how models being trained perform on unseen data, we implemented a validation loop within each epoch where our model runs on our OtterDataset testing set. During validation, we calculate loss using the same loss strategy as the training loop. However, no model parameter updates are done based on this loss; it is strictly for monitoring model performance on unseen data.

\section{Experimental Settings}
\label{sec:settings}

\subsection{Model and Tokenizer Settings}
\label{sub:Encoder model and tokenizer settings}

We trained two variants of our prediction model using the CodeT5 encoder family: CodeT5-small (60 million parameters) and CodeT5-base (220 million parameters). This allowed us to explore how model scale affects cell-level bug detection performance. Additionally, because \textbf{CodeT5 was not pre-trained on Jupyter Notebooks}, there is no data contamination between our training sets and the data used to train CodeT5.

Input tokenization was performed using our custom tokenization strategy in conjunction with the \texttt{RobertaTokenizer} associated with each encoder. While the original CodeT5 models were pre-trained on sequences up to 512 tokens, we increased the maximum sequence length to 2,500 tokens to accommodate longer notebooks and preserve cell-level context.

\subsection{Training Settings}

\label{sub:Training strategy settings}
During training, we used up to the first 4 chunks from each notebook, which allowed us to include all code cells from 19,728 of the notebooks in our dataset (93\%). We considered raising the number of chunks, but didn’t because training on additional chunks would increase GPU memory consumption and per-epoch training time (already 2.09 hours per epoch for JupOtter-base on an A100). This isn’t a limitation of the technique, as we were able to tokenize more than 4 chunks; it is a hardware constraint during training.

We used a batch size of 4 and 10 epochs. A random seed of 42 was used to split our dataset into a training (80\% of total data, 46.6\% buggy) and a testing set (20\% of total data, 49\% buggy).
Optimization was performed using the \texttt{AdamW} optimizer with a learning rate of 5e-5. We enabled mixed precision training to improve training speed and reduce GPU memory consumption.
Training was conducted on an NVIDIA V100 (32GB) GPU for JupOtter-small and on an NVIDIA A100 GPU (40GB) for JupOtter-base. Training time was 2.09 hours per epoch for JupOtter-base, and 2.15 hours for JupOtter-small.

\begin{table}[tbp]
\centering
\caption{Summary of datasets used for evaluation after tokenization.}
\begin{tabular}{@{}l|rr|rr@{}}
\toprule

\textbf{Dataset} & \textbf{\# Files} & \textbf{\# Buggy files} & \textbf{\# Cells} & \textbf{\# Buggy Cells} \\
\midrule
OtterDataset Test & 4,113 & 2,006 (49\%) & 124,364 & 8,598 (7\%) \\
Jupyter Errors & 9,313 & 8,740 (94\%) & 282,371 & 15,960 (6\%) \\
CodeParrot & 4,769  & 1,108 (23\%) & 78,313 & 3,797 (5\%) \\
\bottomrule
\end{tabular}

\label{tab:dataset-summary}
\end{table}

\subsection{Evaluation Benchmark Settings}
\label{sub:External Dataset Collection}

To evaluate the generalization ability of JupOtter beyond its training distribution, we curated and re-labeled two external datasets of Jupyter Notebooks: the Jupyter Errors Dataset~\cite{grotov2024untangling} and a subset of the CodeParrot GitHub Jupyter dataset~\cite{codeparrotgithubjupyter}, both publicly available on Hugging Face. Because no existing benchmarks offered cell-level bug labels, we applied the same annotation strategy used for our OtterDataset to enable consistent evaluation. Table~\ref{tab:dataset-summary} displays a summary of our three evaluation datasets after tokenization. Files are considered buggy if they contain an error within the first four 2,500 token chunks. Using the author, file, and repository name of each file, no overlap was found in the datasets.

\subsubsection{Jupyter Errors Dataset}
The Jupyter Errors Dataset contains 10,000 real-world Jupyter Notebooks, each known to contain at least one runtime error~\cite{grotov2024untangling}.
We labeled the notebook cells in this dataset as buggy or not-buggy, similarly to OtterDataset's labeling process. All notebooks were saved with a file name comprised of the author's name, repository name, and original file name. 660 notebooks were not included due to having duplicate author, file, and repository names, and another 19 were removed due to structural issues, leaving us with a final testing set of 9,321 notebooks. During tokenization, a further 8 were removed due to containing only cells that exceeded our chunk length limit.

\subsubsection{CodeParrot dataset}

The CodeParrot GitHub Jupyter contains 164,619 Jupyter Notebooks collected from GitHub ~\cite{codeparrotgithubjupyter}. 
For our evaluation, we selected the first 5,000 valid Python notebooks from this dataset. 
Similar to our other datasets, notebooks were saved with a file name comprised of the author's name, repository name, and file name found on GitHub. A total of 108 notebooks were excluded due to duplicate author, file, and repository names, resulting in a final testing set of 4,892 notebooks. A further 123 notebooks were removed during tokenization due to token and chunk size constraints. We labeled this subset using the same strategy applied to the previous datasets.

\section{Evaluation \& Results}
\label{sec:results}

We evaluate JupOtter using three datasets: a held-out portion of our OtterDataset, a 4,769-notebook subset of CodeParrot ~\cite{codeparrotgithubjupyter}, and the Jupyter Errors Dataset (9,313 notebooks)~\cite{grotov2024untangling}. We compare JupOtter against Flake8, GPT-4o-mini, and Gemini 3 Flash in RQ1 and RQ3 where we evaluate cell-level and file-level defect prediction. We further evaluate JupOtter in RQ2 where we explore its ability to detect specific bug types, and perform an ablation study in RQ4.

\subsection{RQ1: How well does JupOtter detect bugs at the cell level?}
\label{sub:Results of cell level bug detection model}

\begin{table*}[tbp]
\centering
\caption{Cell-level results for JupOtter, Flake8, GPT-4o-mini, and Gemini 3 Flash on different benchmarks. Best results are in bold.}

\begin{tabular}{l|rrrr|rrrr|rrrr}
\toprule
\multicolumn{1}{c|}{\multirow{2}{*}{\textbf{Tech.}}} &
\multicolumn{4}{c|}{\textbf{OtterDataset Test}} &
\multicolumn{4}{c|}{\textbf{CodeParrot}} &
\multicolumn{4}{c}{\textbf{Jupyter Errors}} \\
 & Pre. & Rec. & F1 & Acc. & Pre. & Rec. & F1 & Acc. & Pre. & Rec. & F1 & Acc. \\
\midrule
\textbf{Flake8} &
0.75 & 0.69 & 0.72 & 0.80 &
0.62 & 0.90 & 0.74 & 0.69 &
0.63 & 0.43 & \textbf{0.51} & 0.69 \\

\textbf{JupOtter$_{small}$} &
\textbf{0.93} & 0.69 & 0.79 & \textbf{0.95} &
\textbf{0.94} & 0.92 & \textbf{0.93} & \textbf{0.97} &
\textbf{0.93} & 0.17 & 0.28 & \textbf{0.91} \\

\textbf{JupOtter$_{base}$} &
0.88 & 0.75 & \textbf{0.81} & \textbf{0.95} &
0.86 & 0.91 & 0.89 & 0.96 &
0.82 & 0.22 & 0.34 & 0.91 \\

\textbf{GPT$_{4o\text{-}mini}$} &
0.26 & 0.85 & 0.40 & 0.74 &
0.16 & \textbf{0.95} & 0.28 & 0.70 &
0.16 & 0.68 & 0.26 & 0.66 \\

\textbf{Gemini$_{3\text{-}Flash}$} &
0.45 & \textbf{0.88} & 0.59 & 0.85 &
0.32 & \textbf{0.95} & 0.48 & 0.78 &
0.32 & 0.72 & \textbf{0.44} & 0.76 \\
\bottomrule
\end{tabular}
\label{tab:rq1}
\end{table*}

This research question evaluates how well JupOtter generalizes to new data, including both held-out notebooks from OtterDataset and two external benchmark datasets.

We evaluate two versions of JupOtter, JupOtter-small and JupOtter-base, trained on CodeT5-small and CodeT5-base, respectively. Both versions are trained on 80\% of OtterDataset. 20\% of the instances are held out and used as a test set. We also evaluate Flake8, GPT-4o-mini, and Gemini 3 Flash, configured for cell-level bug detection, to establish a baseline of performance. We use three test sets for evaluation: OtterDataset Test Set, CodeParrot GitHub Jupyter subset~\cite{codeparrotgithubjupyter}, and the Jupyter Errors Dataset~\cite{grotov2024untangling}.

We report both cell-aggregated and file-aggregated metrics (F1, precision, recall, accuracy). Accuracy figures are included in results tables for completeness and should be interpreted with caution given the class imbalance (approximately 7\% buggy cell rate). In cell-aggregated evaluation, we accumulated true positives, false positives, true negatives, and false negatives across all notebook cells in the test set to compute overall precision, recall, F1 score, and accuracy. 
This metric does not distinguish between notebooks, each cell being treated as an independent unit. This method gives larger notebooks with more cells, more influence on the evaluation metrics since each cell contributes equally to the final evaluation metrics. These cell-aggregated metrics measure how many cells are correctly identified by our model.

The file-aggregated metrics are computed for each notebook in our testing set individually. We then averaged these metrics across all notebooks and used the averaged precision and recall to compute a final averaged F1 score.
These file-aggregated metrics reduce the potential bias created by large notebooks. 
Since \textbf{RQ1} focuses on cell-level performance across multiple files and datasets, we only report file-aggregated metrics in this paper, while cell-aggregated results are available in supplementary material~\cite{jupotter_repo}.

Flake8 was adapted for cell-level bug detection by first mapping each notebook cell to its corresponding line numbers, preserving the line numbering across cells. The notebooks were then converted into Python scripts, allowing Flake8 to analyze the entire file. We considered running Flake8 on each cell individually, but this approach caused many false positives. We also only consider errors raised by Flake8 and ignore warnings and style suggestions. When Flake8 detected an error, we mapped the error’s line number back to the corresponding cell and created a prediction array where the buggy cell was marked with a 1, and all preceding cells were marked with 0. Since Flake8 only reports up to the first fatal error it encounters without reporting subsequent errors, subsequent cells were not considered. We trimmed the labels to match the number of cells analyzed by Flake8. If no errors were detected, the prediction array consisted entirely of 0s, indicating no buggy cells.

Additionally, we configured GPT-4o-mini and Gemini 3 Flash for cell-level bug detection. Each notebook was presented to the models incrementally: at each step, the models were given the current cell together with all previously seen cells and generated predictions, and a prediction was recorded for the current cell. Thus, the models precisely labeled each cell while having access to the full prefix of the notebook.

Due to the large number of API calls required to generate predictions for all cells (485,048), we used random sampling and selected 100 notebooks from each dataset (7,628 cells, 6\% cell-level bug rate) to construct a manageable evaluation set for Gemini 3 Flash and GPT-4o-mini. Gemini 3 Flash averaged 4.8 minutes per notebook, making evaluation on the full test set impractical. The initial prompt was: ``You are analyzing a Jupyter Notebook cell-by-cell. You must remember previous cells. For each cell, answer ONLY with YES or NO indicating whether THIS CELL contains a bug.'' Each subsequent cell and the model’s previous responses were appended to this prompt, forming the incremental prompting scheme.

We only accepted outputs that matched exactly YES or NO after removing punctuation, white spaces, and converting characters to uppercase. 
Once a model produced an output that did not conform to this format, evaluation of the current notebook was terminated. Notebooks for which the model failed to produce a complete sequence of valid predictions were evaluated by truncating the label sequence to the number of cells successfully processed. Gemini 3 Flash was able to generate predictions for 7,626 cells (99.99\%), while GPT-4o-mini was able to produce predictions for 7,561 cells (99.12\%).

Table~\ref{tab:rq1} summarizes results for JupOtter-base, JupOtter-small, Flake8, GPT-4o-mini, and Gemini 3 Flash when evaluated across all three datasets. 
On the held-out OtterDataset test set, JupOtter-base achieves a precision, recall, and F1 of 0.88, 0.75, and 0.81, respectively, while JupOtter-small achieves 0.93, 0.69, and 0.79. 
JupOtter-base and JupOtter-small achieved their best cell-level bug detection performance on our labeled subset of the CodeParrot GitHub Jupyter dataset, with JupOtter-base achieving an F1 score of 0.89 and an accuracy of 0.96, and JupOtter-small achieving an F1 score of 0.93 and an accuracy of 0.97. 
Their strong performance on both datasets suggests JupOtter is able to generalize well and did not overfit to its training set.

However, on the Jupyter Errors dataset, JupOtter maintains high precision but achieves a lower F1 score, indicating that it misses a subset of runtime errors rather than over-predicting bugs. This may be due to the limited quantity of some runtime errors in the OtterDataset, which we used for training, when compared to other error types. Additionally, cells containing runtime bugs that were never executed are labeled non-buggy in our ground truth, meaning the training signal for runtime errors is inherently partial. Consequently, reported recall figures should be interpreted as lower bounds on true performance. RQ2 investigates in more detail which types of error JupOtter can detect well and which ones it struggles with.

We measured the statistical significance of the performance difference between the two JupOtter models and found them significant (Wilcoxon Signed-Rank Test, p $<$ 0.01 on all three benchmarks). This was not done between Flake8, GPT-4o-mini, and Gemini 3 Flash due to notebooks for which they were unable to generate complete predictions.

Moving from a smaller encoder (CodeT5-small, 60 million parameters) to a larger one (CodeT5-base, 220 million parameters) yielded only marginal improvements in two benchmarks (OtterDataset Test and Jupyter Errors) while slightly decreasing performance in CodeParrot. The cost trade-off of JupOtter-base and JupOtter-small is discussed in Section~\ref{sub:Training strategy settings}.

JupOtter demonstrates strong generalization to new notebooks, particularly those containing typical implementation errors. Its performance drops slightly on notebooks dominated by rarer runtime errors, which may be addressed in future dataset augmentation.
These results suggest that JupOtter can be deployed in real-world notebook development environments to flag buggy cells with high accuracy, particularly when errors follow common Python patterns.

We additionally analyzed whether JupOtter’s predictions are influenced by cell size rather than code content by measuring token counts per cell in OtterDataset. We compared the token distributions of cells predicted as buggy and non-buggy using both Cliff’s Delta and the Mann–Whitney U test. While the difference was statistically significant (p $<$ 0.001), the effect size was negligible (Cliff’s Delta = 0.081), and the difference in median token count between the two groups was only 4 tokens. This indicates that cell length alone does not meaningfully explain JupOtter’s predictions. This observation is consistent with prior work showing that bugs in notebooks are primarily correlated with developer behavior rather than superficial code properties~\cite{jiang2025exploring}.

\smallskip\finding{\textbf{RQ1 Summary:} JupOtter achieves strong cell-level bug detection performance on multiple unseen datasets, with \textbf{best F1 of 0.93 and accuracy of 0.97}, indicating good generalization.}

\subsection{RQ2: How does JupOtter perform when detecting specific error types?}
\label{sub:specific-error-eval}

This research question evaluates whether JupOtter’s performance varies across different categories of bugs, and whether certain error types are easier or harder for the model to detect.

We selected the five most frequent error categories from OtterDataset: TypeError, NameError, SyntaxError, ValueError, and Attri\-buteError. For each category, we evaluated JupOtter-base on notebooks containing at least one cell with the corresponding error. These notebooks were drawn from the labeled subset of the CodeParrot GitHub Jupyter dataset to ensure that the evaluation was performed on external data.

\begin{table}[ht]
\caption{Cell-level performances of JupOtter-base on the five most common error types.}
\centering
\begin{tabular}{@{}l|c|c|c|c@{}}
 \toprule
 \textbf{Error Type} & \textbf{Recall} & \textbf{Precision} & \textbf{F1} & \textbf{Accuracy} \\ 
 \midrule
 Attribute errors & 0.28 & 0.63 & 0.38 & 0.92 \\

 Name errors & 0.23 & 0.65 & 0.34 & 0.92 \\ 

 Syntax errors & 0.85 & 0.90 & 0.87 & 0.96 \\ 

 Type errors & 0.51 & 0.84 & 0.64 & 0.94 \\ 
 
 Value errors & 0.51 & 0.82 & 0.63 & 0.95 \\
 
 \bottomrule
\end{tabular}
\label{tab:specific-error-eval}
\end{table}

Table~\ref{tab:specific-error-eval} shows the cell-aggregated results of JupOtter-base across the five error types. JupOtter achieves the strongest performance (F1: 0.87) for SyntaxError; performs moderately well on TypeError and ValueError with an F1 score of 0.64 and 0.63, respectively; and struggles on AttributeError (F1: 0.38) and NameError (F1: 0.34).

\smallskip\finding{\textbf{RQ2 Summary:} JupOtter performed well for Type, Value, and Syntax errors, but struggled for Attribute and Name errors. High accuracy is maintained in all cases, but recall is significantly reduced for hard-to-catch error types.}

\subsection{RQ3: How does JupOtter compare to related tools for file-level bug detection?}
\label{sub:File level bug detection}

RQ3 evaluates JupOtter’s performance on file-level bug detection tasks and compares it to traditional static analysis tools and prior work using language models.

Since no prior work has focused on cell-level bug detection, we cannot directly compare JupOtter to existing cell-level approaches. To evaluate our baselines in a setting most similar to related tools, we evaluate JupOtter’s file-level bug detection ability, enabling a direct comparison with Flake8, GPT-4o-mini, and Gemini 3 Flash. This file-level evaluation provides a more direct comparison to JupOtter and related work than the cell-level analysis in Section~\ref{sub:Results of cell level bug detection model}, where baselines were adapted for cell-level analysis. Because file-level detection does not require multiple API calls per notebook, whereas cell-level detection requires a separate API call for each cell, GPT-4o-mini and Gemini 3 Flash can be evaluated on the entirety of the test datasets, enabling a substantially larger comparison without random sampling.

We considered comparing JupOtter against existing file-level defect prediction models, such as DeepLineDP~\cite{pornprasit2022deeplinedp}, and LineVul~\cite{fu2022linevul}, but did not include them in our evaluation. Existing defect prediction models are designed for traditional source code files and are not trained to process Jupyter Notebook-specific syntax, dynamic execution, or cell structure. Including existing defect prediction models would unfairly showcase baselines outside their intended domain. The large language models we evaluate are general-purpose, and Flake8 can be configured to analyze notebook code, making these baselines more appropriate for comparison.

To evaluate Flake8, we transformed notebook files into Python scripts since Flake8 doesn't work directly on notebooks. Using GPT-4o-mini and Gemini 3 Flash for bug detection at scale requires automatically scanning model outputs and determining if a potential bug has been flagged. To achieve this, we prompted the models to respond with only binary answers (``YES'' or ``NO'') using the instruction: ``You are performing file-level bug detection. Respond YES if there is a bug in the file; otherwise, respond NO. Only ever respond with YES or NO.'' Capitalization and punctuation in the model outputs were removed for consistency. Out of 18,195 answers generated by GPT-4o-mini, 1,055 (6\%) were non-binary, while Gemini 3 Flash produced 3,646 (20\%) non-binary answers. These responses were omitted from our evaluation, as their large number made manual verification impractical.

\begin{table*}[ht]
\centering
\caption{File-level results for JupOtter-base, Flake8, GPT-4o-mini, and Gemini 3 Flash on different benchmarks. Best results are in bold.}
\begin{tabular}{l|rrrr|rrrr|rrrr}
\toprule
\multicolumn{1}{c|}{\multirow{2}{*}{\textbf{Tech.}}} &
\multicolumn{4}{c|}{\textbf{OtterDataset Test}} &
\multicolumn{4}{c|}{\textbf{CodeParrot}} &
\multicolumn{4}{c}{\textbf{Jupyter Errors}} \\
 & Pre. & Rec. & F1 & Acc. & Pre. & Rec. & F1 & Acc. & Pre. & Rec. & F1 & Acc. \\
\midrule
\textbf{Flake8} & 0.61 & 0.77 & 0.68 & 0.64 & 0.31 & \textbf{0.94} & 0.47 & 0.51 & 0.92 & 0.72 & 0.81 & 0.69 \\
\textbf{JupOtter$_{base}$} & \textbf{0.88} & 0.67 & \textbf{0.76} & \textbf{0.80} & \textbf{0.62} & 0.76 & \textbf{0.68} & \textbf{0.83} & \textbf{0.94} & 0.35 & 0.51 & 0.38 \\
\textbf{GPT$_{4o\text{-}mini}$} & 0.55 & 0.87 & 0.67 & 0.58 & 0.29 & 0.91 & 0.44 & 0.46 & 0.92 & \textbf{0.83} & \textbf{0.87} & \textbf{0.78} \\
\textbf{Gemini$_{3\text{-}Flash}$} & 0.61 & \textbf{0.88} & 0.72 & 0.66 & 0.33 & 0.87 & 0.48 & 0.58 & 0.93 & 0.78 & 0.85 & 0.75 \\
\bottomrule
\end{tabular}
\label{table:file_results_comparison}
\end{table*}

Table~\ref{table:file_results_comparison} reports the F1 scores, precision, and accuracy of file-level predictions. JupOtter-base outperformed GPT-4o-mini and Gemini 3 Flash on two of the three test datasets in terms of F1 score. JupOtter achieved F1 scores of 0.76 and 0.68 on the OtterDataset test set and CodeParrot respectively, while Gemini 3 Flash had an F1 score of 0.72 on OtterDataset test set, and all 3 baselines achieved an F1 score of 0.48 or less on the CodeParrot dataset. GPT-4o-mini performed best on the Jupyter Errors dataset, with an F1 score of 0.87.

GPT-4o-mini, Gemini 3 Flash, and Flake8 perform strongly on the Jupyter Errors dataset, where every untruncated notebook contains at least one runtime error (94\% of notebooks in our truncated testing set). This strong performance on the Jupyter Errors dataset is likely due to the baselines being biased towards classifying notebooks as buggy, shown by their lower precision on the two other datasets.

These results indicate that JupOtter complements existing static analysis and language model-based tools. While Flake8, Gemini 3 Flash, and GPT-4o-mini tend to classify notebooks as buggy, resulting in higher false positive rates, JupOtter-base is more conservative, with higher precision and performing strongly on datasets containing error-free notebooks, such as OtterDataset and CodeParrot. This suggests JupOtter may be particularly useful in file-level bug detection scenarios where minimizing false positives is important, whereas Flake8, Gemini 3 Flash, or GPT-4o-mini may be more suitable when maximizing recall is the priority.

The weakness of JupOtter-base on the Jupyter Error benchmark may suggest that it struggles when detecting runtime errors present in this dataset. Overall, these results suggest that although file-level bug detection is not the intended use for JupOtter-base, it still outperforms traditional static analysis tools and general large language models at file-level bug detection tasks, though runtime errors remain a challenge.

To measure whether there is a statistically significant difference in the performance of JupOtter-base and Flake8, we used the McNemar test, measuring the significance of the difference in performance on each dataset. We found a significant statistical difference with p values $<$ 0.01 on all three testing sets. We did not perform paired statistical testing between GPT-4o-mini, Gemini 3 Flash, and JupOtter because GPT-4o-mini and Gemini 3 Flash were unable to generate predictions for every notebook, leaving many samples unpaired.

\smallskip\finding{\textbf{RQ3 Summary:} {JupOtter-base outperformed the baselines in terms of F1 score on two of three test datasets, achieving \textbf{F1 scores of 0.76 vs. 0.72 and 0.68 vs. 0.48} relative to the strongest performing baseline.}}

\subsection{RQ4: How does each contribution affect performance?}
\label{sub:ablation}

In this section, we present an ablation study to quantify the contribution of JupOtter’s components, in particular our notebook-aware tokenization strategy and our use of multi-segment processing for long notebooks. Models in this section were tested using the OtterDataset test set.

\subsubsection{Impact of the new tokenization strategy}
Since CodeT5 cannot perform bug detection out of the box, we first trained a baseline model using the original CodeT5 tokenizer with a classification head that outputs a single logit for file-level prediction. This model achieves an F1 score of 0.73, compared to 0.76 for JupOtter-base. More importantly, this minimal baseline cannot localize bugs at the cell level and fails to fully tokenize 84\% of notebooks due to context length limitations, compared to only 7\% with JupOtter using our tokenization strategy. This demonstrates that our new tokenization strategy is not only necessary for localization but also crucial for processing the majority of notebooks.

While our tokenization strategy has a limited impact on file-level prediction (+3\% F1), it is required for predictions at the cell-level, as it enables distinguishing between individual cells and allows processing of 77\% of notebooks that would otherwise exceed token limits.

Additionally, we attempted to train a model that directly classifies token embeddings without first using an encoder. This model failed to learn and predicted all samples as non-buggy (the majority class), confirming that contextual encoding is required prior to classification.

\subsubsection{Impact of the multi-segment tokenization}
We trained a cell-level model that uses special boundary tokens to mark cells, but without splitting long notebooks into multiple chunks. This model achieves an F1 score of 0.78, compared to 0.81 for the full JupOtter model. While the performance drop is moderate, this baseline suffers from the same problem as our minimal baseline: it is unable to fully process 84\% of notebooks due to context length limitations. In contrast, JupOtter’s tokenization strategy allows the model to scale to much larger notebooks using the same encoder size, where each chunk used effectively adds an entire encoder’s worth of tokens which can be processed. 

\smallskip
\finding{\textbf{RQ4 Summary:} While JupOtter’s components provide performance improvements, their primary benefit is a substantial increase in scalability to large notebook lengths, while also enabling precise cell-level localization.}

\section{Threats to Validity}
\label{sec:limits}

\noindent\textbf{Data Leakage:} A potential threat to the validity of our results is data leakage, specifically, the possibility that the models may have access to indirect information about the bugs under evaluation. We mitigate this threat in three ways.  First, the pre-trained model we use, CodeT5, was trained on the CodeSearchNet dataset~\cite{husain2019codesearchnet} and additional data from C/C\# repositories, none of which include Jupyter Notebooks. Therefore, there is no overlap between the pre-training corpus and our evaluation data. Second, for fine-tuning, we ensured that all duplicate files were removed to prevent any instance of the same notebook appearing in both the training and testing subsets of OtterDataset. Third, we addressed a common threat in defect prediction studies, where a test set includes buggy files whose fixed versions appear in the training set. This can introduce unrealistic ``future knowledge'' into the model. To avoid this, we did not use version history or commit data during the construction of OtterDataset, and ensured that multiple revisions of the same notebook could not appear across training and test sets. We cannot guarantee the absence of data leakage for the Gemini 3 Flash and GPT-4o-mini baselines. However, any such leakage would advantage these baselines rather than affect the performance of JupOtter.

\smallskip\noindent\textbf{Generalization of our Results:} The scope of our dataset may limit generalization. We trained and evaluated the model on a specific collection of notebooks that may not fully capture the diversity of coding styles, bug types, or domains. To mitigate this, we collected notebooks using a variety of queries to target diverse domains and error types and used two additional external test sets collected by other researchers.

\smallskip\noindent\textbf{Incorrect Labeling:} The quality of our labeled dataset depends on the accuracy and consistency of the bug annotations, which were derived through automated inspection. Any errors, inconsistencies, or biases in labeling could affect the model’s training and evaluation, potentially leading to underperformance on certain bug types. To assess the quality of our automatic labelling, we manually labeled a sample of our data and only observed minimal inconsistency (see section \ref{subsub:notebook error annotation}).

\section{Related Work}
\label{sec:related}

\smallskip\noindent\textbf{Bugs in Jupyter Notebooks:}
Prior work examined Jupyter Notebooks and their associated bugs on GitHub and Stack Overflow. 
Researchers found that the most common types of bugs are environment errors and implementation errors ~\cite{de2024bug}. 
Environment errors account for 35.6\% of .ipynb errors on GitHub and 43.2\% on Stack Overflow, while implementation errors make up 44.2\% of errors on GitHub and 22\% on Stack Overflow. Additionally, both error types are growing rapidly on Stack Overflow, with implementation errors increasing at a rate of 48\% per year and environment errors at 38\%.

Prior research has analyzed bugs and software across various contexts~\cite{tang2025characterising, pimentel2019large, wang2025machine, tan2014bug, cotroneo2016bugs, wang2020better}. Notably, Wang et al.~\cite{wang2020better} focus specifically on Jupyter Notebooks and highlight that although they are widely adopted by students, researchers, and data scientists, they often suffer from poor coding practices. Their analysis found common issues such as unused variables, lack of adherence to Python style conventions, and inconsistent code structure. The authors argue that these issues reduce the reproducibility and quality of research and advocate for automated tools to improve code quality in notebooks. Our work builds on this motivation by developing an automated technique to detect implementation bugs in Jupyter Notebooks, addressing an aspect of notebook reliability that goes beyond style and formatting issues. 

\smallskip\noindent\textbf{Other Notebook Studies:}
A growing body of research has explored the structure, quality, and usage patterns of Jupyter Notebooks. Several studies investigated the notebook ecosystem and detailed notebook characteristics~\cite{kallen2020jupyter}, common cleaning activities~\cite{dong2021splitting,cunha2021context}, code reuse~\cite{ritta2022reusing}, bugs~\cite{patra2022nalin}, reproducibility~\cite{wang2020assessing,wang2020restoring}, or proposed support for debugging~\cite{merino2022making}.
Others compared notebooks with traditional Python scripts~\cite{grotov2022large,adams2023comparison} or detailed best practices~\cite{quaranta2022eliciting} and challenges in the notebook ecosystem~\cite{settewong2022visualize,chattopadhyay2020s}. Machine learning techniques have also been explored for various tasks in notebook environments such as code generation~\cite{chandel2022training, yin2023natural}. 

\smallskip\noindent\textbf{Defect Prediction and Bug Detection:}
To our knowledge, no prior work has explored cell-level bug detection in Jupyter Notebooks using language models.
Previous work~\cite{wang2024using} is the closest related work and investigates file-level bug detection using GPT-4. The authors synthetically generated notebooks by splitting Python scripts into cells, then applied GPT-4 as a static analyzer to detect bugs, with and without runtime information. While informative, their approach does not operate at the cell-level and lacks real-world notebook structure.

Outside of notebooks, numerous studies have explored defect prediction in traditional source code~\cite{gaddamwar2024deep, jiang2024repaircat, azizi2024astrobug, li2018vuldeepecker, pradel2018deepbugs}. 
ChatDBG~\cite{levin2025chatdbg} integrates LLMs with traditional debuggers to improve the usability and capabilities of traditional debuggers, and LOVA~\cite{li2024if} leverages self-attention mechanisms within LLMs to localize vulnerabilities in source code.
BUGLAB~\cite{allamanis2021self} presents a self-supervised learning approach for bug detection and repair, while DefectHunter~\cite{wang2023defecthunter} employs the Conformer architecture to identify vulnerabilities. Toggle~\cite{hossain2024deep} uses LLMs to predict bugs at the token level. 

Due to the lack of ML baselines for cell-level detection, we evaluated JupOtter against a static analyzer. Many Python static analyzers exist~\cite{quaranta2022pynblint,pylint,pyflakes, sonarqube,subotic2022static}. Among these, Flake8~\cite{flake8} was selected as a baseline for its speed and configurability.

Defect prediction approaches have used semantic representations to improve prediction \cite{wang2018deep}. Additionally, work has evaluated the consistency and reliability of model‑agnostic explainable techniques across different defect prediction settings, emphasizing that explainability remains a challenge even for well‑trained models \cite{shin2021explainable}. Prior work has also highlighted that most benchmark datasets for software bug prediction use randomly selected historical versions, which may not reflect real-world continuous software development \cite{wang2021continuous}.

\smallskip\noindent\textbf{Machine Learning Models for Code:}
The encoder we use to build JupOtter is CodeT5, a Transformer model pre-trained on large-scale code datasets for program understanding tasks~\cite{wang2021codet5}. We chose it as a relatively lightweight model compared to more recent code language models. Its follow-up, CodeT5+, has demonstrated improved performance across benchmarks~\cite{wang2023codet5+}. Other pre-trained models in this space include CodeBERT~\cite{feng2020codebert}, CodeLlama ~\cite{roziere2023code}, GraphCodeBERT~\cite{guo2020graphcodebert}, CuBERT~\cite{kanade2020learning}, and PLBART~\cite{ahmad2021unified}. These models were pre-trained on large amounts of code text data for code understanding tasks such as code summarization, generation, and defect prediction. While we chose CodeT5 as our base model, our approach is not bound to a specific model and could be applied to many pre-trained code models.

We also drew on previous work~\cite{liu2022leveraging} to address the challenge of processing large notebooks. This work processes full source documents as a series of non-overlapping pages rather than as a continuous sequence. Following this idea, we tokenize notebooks into multiple non-overlapping cell sequences.

\smallskip\noindent\textbf{Notebook Datasets:}
We found no existing datasets with cell-level bug annotations for Python Jupyter Notebooks, which motivated the creation of OtterDataset (detailed in Section~\ref{sub:extraction}). While prior datasets~\cite{grotov2024untangling,codeparrotgithubjupyter,mostafavi2024distilkaggle,quaranta2021kgtorrent,chandel2022training,vikp_clean_notebooks_filtered} exist for analyzing notebooks, most are unlabeled or focus on usage patterns and code quality rather than fine-grained bug detection. We also re-labeled two datasets introduced in previous work that included buggy notebooks~\cite{grotov2024untangling,codeparrotgithubjupyter}. 
\section{Implications and Future Work}
\label{sec:discussion}

\noindent\textbf{Practical Deployment:} JupOtter is designed with practical deployment in mind. By supporting long notebooks and generating cell-level predictions, it offers a finer level of granularity than traditional file-level or function-level bug detectors. This enables developers to pinpoint specific faulty cells rather than being presented with vague file-wide warnings. This capability is especially valuable in educational and research contexts, where feedback needs to be actionable and precise.
In contrast, function- or file-level tools may fail to provide applicable predictions for notebooks. By offering predictions at the cell-level, JupOtter better aligns with real-world notebook usage and has the potential to improve both developer productivity and code quality. Additionally, JupOtter’s prediction system can be modified to output a vector of predictions per cell rather than a single logit, enabling the model to indicate not only whether a cell is buggy, but also the type of error present. However, this would require a more extensive labeling process.

A promising direction for future work is integrating JupOtter into popular notebook environments such as JupyterLab or VS Code as a live assistant. This would allow users to receive real-time bug predictions while writing code, much like linting tools for scripts. To support this, future research could explore optimizing the model for latency and memory usage, investigating incremental inference strategies (e.g., predicting only changed cells), and providing confidence scores or explanations alongside predictions. Deployment pipelines could also include automatic triaging of predictions (e.g., flagging only high-confidence bugs) to reduce noise and improve user trust. By addressing these engineering and UX challenges, JupOtter could serve as a practical assistant that helps developers catch errors earlier and code with greater confidence.

\smallskip\noindent\textbf{Extension to Markdown Cells:}
While effective, our tokenization strategy only considers code cells and treats chunking strictly in terms of token length, which excludes context from markdown cells. One future direction is to extend our tokenization strategy to create chunks of tokens based on context boundaries indicated within markdown headings. Incorporating this structure may help models better understand code in context and improve detection.

\smallskip\noindent\textbf{Summary:}
More broadly, JupOtter opens the door to cell-level predictions in notebook environments. Notebooks are widely used in education, research, and data science, yet tools for analyzing and improving their reliability are still limited. Our work provides a foundation for future tools that integrate with notebook workflows and help authors write more correct and maintainable code.

\section{Conclusion}
\label{sec:conclusion}
In this paper, we introduced \textbf{JupOtter}, a machine learning-based system designed specifically for cell-level bug detection in Jupyter Notebooks. Our approach incorporates a novel tokenization strategy that preserves notebook structure, a prediction method that operates at the cell-level, and a training strategy tailored for multi-segment inputs.
We also released \textbf{OtterDataset}, a large-scale labeled dataset for supervised bug detection in Jupyter Notebooks, comprising 21,303 notebook files. Through extensive evaluation across three benchmarks, we demonstrated that JupOtter achieves strong performance on both cell-level and file-level bug detection tasks, outperforming static analysis tools as well as large language models.

\section*{Data Availability}
Training and evaluation code, trained model parameters, configuration for baselines, error examples, datasets, and notebook labeling code are available on Zenodo~\cite{jupotter_repo}. 

\section*{Acknowledgments}
ChatGPT was used for table formatting, text editing (e.g., grammar and clarity), and writing routine portions of code. AI was not used to generate results, figures, datasets, or core contributions. This work was supported by the Natural Sciences and Engineering Research Council of Canada (NSERC).

\bibliographystyle{IEEEtran}
\bibliography{paper}
\end{document}